\documentclass[RNAAS]{aastex62}
\usepackage{amsmath}
\usepackage{bm}
\usepackage{relsize,exscale}

\begin{document}

\title{Note on the Power-2 Limb Darkening Law}

\correspondingauthor{Gur Windmiller}
\email{gwindmiller@sdsu.edu}

\author{Donald R.\ Short}
\affiliation{Department of Astronomy, San Diego State University, 5500 Campanile 
Drive,  San Diego CA 92182, USA}

\author[0000-0003-2381-5301]{William F.\ Welsh}
\affiliation{Department of Astronomy, San Diego State University, 5500 Campanile 
Drive,  San Diego CA 92182, USA}

\author{Jerome A.\ Orosz}
\affiliation{Department of Astronomy, San Diego State University, 5500 Campanile 
Drive,  San Diego CA 92182, USA}

\author{Gur Windmiller}
\affiliation{Department of Astronomy, San Diego State University, 5500 Campanile 
Drive,  San Diego CA 92182, USA}

\author[0000-0003-3794-1317]{P.\ F.\ L.\ Maxted}
\affiliation{Astrophysics Group, Keele University,
Staffordshire, ST5 5BG UK}




\section{} 

Limb darkening laws are simple formulae that approximate the stellar intensity 
$I_{\lambda}$ as function of the foreshortening angle $\mu$ measured from the
center of the stellar disk (i.e.\ $\mu \equiv \cos{\theta}$ where $\theta$ is the 
angle between the line of sight and the surface normal). Recently there has been 
a renewed interest in the power-2 law for modeling exoplanet transits 
(e.g.\ \citet{Maxted2018}, \citet{Maxted2019}) because it provides a better 
match to the stellar intensities generated by spherical stellar atmosphere models 
than other 2-parameter laws \citep{Morello2017}. 
Accuracy in modeling the limb darkening is particularly important when 
attempting to measure unbiased exoplanetary radii 
(e.g. \citet{Espinoza2015}, \citet{Neilson2017})
and higher-order effects such as tidal deformation
\citep{Akinsanmi2019}.
We agree with the above cited works and so to help facilitate wider use 
of the power-2 law, 
we correct a minor error and expand on the work of \citet{Maxted2018}.\\

In the power-2 law (see \citet{Hestroffer1997}) the specific intensity, $I$, 
is defined as 
\begin{align}\label{eq:Iofmu_def}
\begin{aligned}
I(\mu) = 1 - c\big(1-\mu^\alpha\big) ~~~~~~~~ \alpha > 0, ~~~ 0 \le \mu \le 1.
\end{aligned}
\end{align} 
\citet{Maxted2018} found that $c$ and $\alpha$ are highly correlated
when modeling transit light curves, so to enable more efficient sampling 
he introduced the parameters 
$h_1 \equiv I$($\mu$=0.5) and $h_2 \equiv h_1 - I$($\mu$=0)
in place of $c$ and $\alpha$.
These are related to $c$ and $\alpha$ by
\begin{align}\label{eq:power2_coeff1}
\begin{aligned}
& h_1 = 1 - c\big(1-2^{-\alpha}\big) \\
& h_2 = c 2^{-\alpha}
\end{aligned}
\end{align}
and the inverse
\begin{align}\label{eq:power2_coeff2}
\begin{aligned}
&      c = 1 - h_1 + h_2 \\
& \alpha = \log_2\bigg(\frac{c}{h_2}\bigg).
\end{aligned}
\end{align}
\citet{Maxted2018} further stated that ``These definitions impose the 
constraints $h_1 < 1$ and $\bf{h_1+h_2 \le 1}$ that are required so that 
the flux is positive at all points on the stellar disc.'' 
Unfortunately, the inequality highlighted in bold does not, in fact, satisfy 
that condition. 
For example, if we choose $h_1 = 0.25$ and $h_2 = 0.50$ this gives 
$I(0) = -0.25$, an unphysical negative intensity at the center 
of the stellar disk.
Below we provide a derivation of the actual realizable 
region in the $(h_1,h_2)$ plane -- a triangular area obeying the inequalities 
$h_1 < 1$ and $0 < h_2 \le h_1$.
Knowing the allowed region of the power-2 limb darkening parameter space is
important for Bayesian parameter estimation techniques, such as the various 
flavors of MCMC, since these methods require proper sampling of the priors. \\

Following \citet{Kipping2013}, the necessary physical constraints on $I$ are
\begin{align}\label{eq:phys_conditions}
\begin{aligned}
&{\rm Condition \ A:} \ I(\mu) \ge 0  \\
&{\rm Condition \ B:} \ I(\mu) \text{ is a strictly increasing function 
(from limb to center)}
\end{aligned}
\end{align} 
Note that conditions A and B imply that the $h_1$ and $h_2$ parameters are 
positive. The function $\mu^\alpha$ is a root for $0 < \alpha < 1$ 
and a power for $\alpha \ge 1$. 
In both cases, $\mu^\alpha$ is an increasing function, and the 
$min_\mu~\mu^\alpha = 0$. 
If $c > 0$ then $c\mu^\alpha$ is strictly increasing; otherwise, 
if $c < 0$ 
then $c\mu^\alpha$ is decreasing. 
Condition B thus implies that $c > 0$. 
Condition A implies
\begin{align}\label{eq:condA_implies}
\begin{aligned}
I(\mu) \ge min_\mu I(\mu)=1-c+min_\mu(c\mu^\alpha)=1-c \ge 0 
~~~ \Rightarrow c \le 1
\end{aligned}
\end{align}
The region in the $(c,\alpha)$ plane is thus the semi-infinite strip 
given by $0 < c \le 1$ and $\alpha > 0$.
The region boundaries in the $(c,\alpha)$ plane then imply a restriction 
in the $(h_1,h_2)$ plane:
\begin{align}\label{eq:restict_plane1}
1 \ge c ~~ \Rightarrow ~~ 1 \ge 1-h_1+h_2 ~~~ \Rightarrow ~~ h_1 \ge h_2
\end{align}
\begin{align}\label{eq:restict_plane2}
\alpha > 0 ~~ \Rightarrow ~~ \log_2{\bigg(\frac{c}{h_2}\bigg)} > 0 
~~ \Rightarrow 
\frac{c}{h_2} > 1 ~~~ \Rightarrow ~~ c > h_2  \\
c > h_2  ~~ \Rightarrow ~~ (1 - h_1 + h_2) > h_2 ~~~ \Rightarrow ~~ 1 > h_1
\end{align}
Thus, the region in the $(h_1,h_2)$ plane is triangular and given by 
the inequalities 
\begin{align}\label{eq:triangle_ineq}
h_1 > 0, ~~ h_2 > 0, ~~h_2 \le h_1, ~~ \text{and} ~ h_1 < 1.
\end{align} \\

\begin{figure}[ht!]\
\plotone{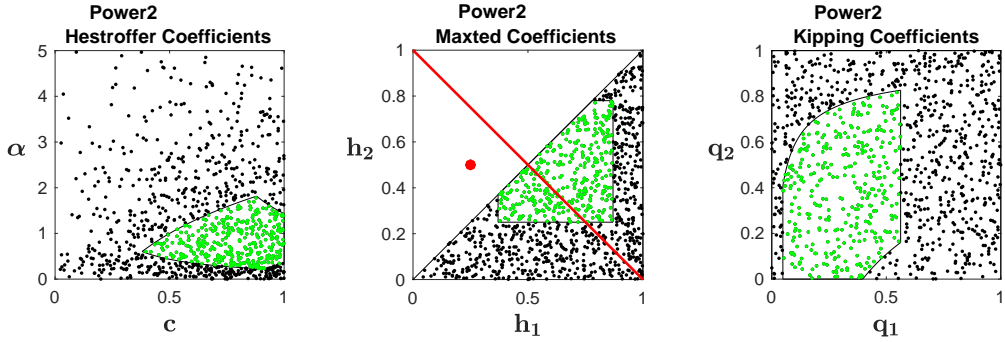}
\caption{The figure shows the limb darkening coefficients for the power-2 law.
The left panel shows a small region of the semi-infinite strip corresponding 
to the Hestroffer $c$ and $\alpha$ coefficients.
The middle panel shows the triangular region using the Maxted $h_1$ and $h_2$ 
coefficients. Note that the red line is an incorrect bounding line segment and  
the red point at ($h_1$=0.25, $h_2$=0.5) resides outside the physically allowed 
region. The right panel shows the ($q_1$, $q_2$) unit square in the 
Kipping parameterization. 
The dots show 1000 uniform samples randomly chosen from the ($q_1$, $q_2$) 
unit square, and then cast into the Maxted and Hestroffer regions.
The green colored dots correspond to the more likely pairs of coefficients, 
based on stellar atmosphere models as found in \citet{Maxted2018}.
}
\label{fig:ThreePanelGraph}
\end{figure}

For MCMC methods, efficient sampling of the prior can be extremely important.
A new set of parameters can be defined which transforms the ($h_1$, $h_2$) 
triangular region into the unit square, preserving the uniform sampling property. 
Such a transformation can be obtained by following the prescription given by 
\citet{Kipping2013}:
\begin{align}\label{eq:prescript_Kipping}
\begin{aligned}
& q_1 = (1-h_2)^2 \\
& q_2 = (h_1-h_2)/(1-h_2) \\
& {\rm for} ~~ 0 < q_1 < 1, ~~~ 0 \le q_2 < 1
\end{aligned}
\end{align}
with the inverse transformation given by:
\begin{align}\label{eq:inverse_trans}
\begin{aligned}
&h_1=1-\sqrt{q_1}+q_2\sqrt{q_1} \\
&h_2=1-\sqrt{q_1}.
\end{aligned}
\end{align}
Finally, the parameters $c$ and $\alpha$ can then be written in terms of 
$q_1$ and $q_2$:
\begin{align}\label{eq:inverse_trans_calpha}
\begin{aligned}
&     c  = 1 - q_2 \sqrt{q_1} \\
& \alpha =        \log_2 \bigg(\frac{1-q_2\sqrt{q_1}}{1-\sqrt{q_1}}\bigg)
         = 3.3219 \log{  \bigg(\frac{1-q_2\sqrt{q_1}}{1-\sqrt{q_1}}\bigg). }
\end{aligned}
\end{align}
Figure \ref{fig:ThreePanelGraph} shows the allowable regions for 
the power-2 law coefficients in the Kipping, Maxted, and original Hestroffer
formulations.
Note that while the triangular region defined by $h_1$ and $h_2$ is permitted 
by the power-2 law parameterization and conditions on $I$, 
not all allowed locations in the region are {\it a priori} equally likely.
Some areas correspond to intensity distributions that may not be
realized in actual stellar atmospheres.
For example, using the 3D {\textsc{stagger}}-grid \citep{Magic2015} 
of stellar atmosphere models for cool stars,
\citet{Maxted2018} finds 
no cases outside the ranges
$0.37 < h_1 < 0.87$ or  
$0.25 < h_2 < 0.78$
for unspotted stars.
This corresponds to a significantly smaller region than the unit square in the 
($q_1$, $q_2$) plane, and a greatly reduced area in the ($c$, $\alpha$) plane.
Such limits can be imposed by placing appropriate constraints on the priors.


\vspace*{0.8cm}

\acknowledgments
We acknowledge support from the NSF via grant AST-1617004, and thank 
John Hood, Jr.\ for his support of exoplanet research at SDSU.

\end{document}